# Universal Swarm Computing by Nanorobots

**Alireza Rowhanimanesh[1,2] and Mohammad-R. Akbarzadeh-T.[2] ***

[1] Department of Electrical Engineering, University of Neyshabur, Neyshabur, Iran.
[2] Department of Electrical Engineering, Center of Excellence on Soft Computing and
Intelligent Information Processing, Ferdowsi University of Mashhad, Mashhad, Iran.
* E-mail: akbazar@um.ac.ir

***Abstract.*** Realization of universal computing units for nanorobots is highly promising in creating new and wide arrays of applications, particularly in the realm of distributed computation. However, such realization is also a challenging problem due to the physical limitations of nanometer-sized designs such as in computation, sensory and perception as well as actuation. This paper proposes a theoretical foundation for solving this problem based on a novel notion of distributed swarm computing by basis agents (BAs). The proposed BA is an abstract model for nanorobots that can compute a very simple basis function called B-function. It is mathematically shown here that a swarm of BAs has the universal function approximation property and can accurately approximate functions. It is then analytically demonstrated that a swarm of BAs can be easily reprogrammed to compute desired functions simply by adjusting the concentrations of BAs in the environment. We further propose a specific structure for BAs which enable them to perform distributed computing such as in the aqueous environment of living tissues and nanomedicine. The hardware complexity of this structure aims to remain low to be more reasonably realizable by today's technology. Finally, the performance of the proposed approach is illustrated by a simulation example.

***Keywords*****:** Nanorobotics, Swarm Computing, Universal Function Approximation, Programmability.

## I. INTRODUCTION

A prerequisite for intellectualizing a nanorobot is equipping it with a nano-sized computing unit to perform computations. However, realization of programmable computers by nanometer-sized agents is a challenging open problem since nanoscale components are highly limited in actuation, computation, communication, as well as mobility. This is an important need for nanomedicine where nanorobots are expected to operate in highly complex and uncertain environments such as the human body when treating cancer, atherosclerosis, and Alzheimer's disease. Furthermore, external programmability is highly desirable in many of these applications, particularly for physicians, to program nanorobot from outside the human body to execute desired function distributions.

The previous experimental works in the literature, which are discussed in the next section, are promising in showing that smaller nanoscale components can be manufactured in near future. However, they all concur that a single nanorobot has very limited capabilities either due to the inherent and physical limitations of this paradigm or due to a need for further technological advancements. In other words, the hardware complexity of any realizable nanorobot should remain as low as possible to facilitate its manufacturing within the existing bounds of technologies. Fortunately, most applications of nanotechnology such as in nanomedicine have access to large swarms of nanoscale agents in even small volumes of the environment. The central question here is: can we exploit a swarm of very simple nanometer-sized agents to perform complex computations in a distributed manner?

There exist several works in the literature that have applied swarm micro/nanorobotics to nanobiomedical applications including drug delivery, microsurgery, in-vivo sensing, and detoxification [1]. Yu et al. [2-3] mathematically analyzed and experimentally studied active generation, magnetic actuation, and pattern formation of swarms of nanoscale agents in bio-fluids. Wang et al. [4-5] proposed a navigation strategy for a nanoparticle microswarm to connect swarm control and real-time imaging for localized drug delivery in dynamic environments. Yigit et al. [6] presented an approach for programmable self-assembly, propulsion, and collective behavior of mobile microrobot swarms based on magnetic interactions. Xie et al. [7] designed a reconfigurable magnetic microrobot swarm as a functional bio-microrobot system for biomedicine with capabilities of multimode transformation, locomotion, and manipulation.

Loscrí and Vegni [8] introduced an acoustic communication paradigm for a swarm of medical nanorobots to perform therapeutical tasks in a distributed and decentralized biological environment. Wang et al. [9] demonstrated that a system of nanoparticles can exploit one of the body's own communication pathways to communicate with one another to raise the efficiency of targeted nanomedicine. Shi et al. [10] developed an in-vivo computation framework for nanorobots-assisted tumor targeted therapy. Rowhanimanesh and Akbarzadeh [11-14] introduced the notion of swarm control systems for nanomedicine based on autonomous drug-encapsulated nanoparticles, and considered its application to the treatment of atherosclerosis through in-silico study. Raz et al. [15-16] designed bioinspired nanonetworks with capability of swarm intelligence for targeted cancer drug delivery.

In addition to the above-mentioned nanoscale applications, many computational frameworks are available in multidisciplinary areas such as swarm intelligence [17] and swarm robotics [18] for distributed problem solving by swarms of simple agents. Here, we review some of the most related references to this paper. Cellular automata [19] as well as several types of artificial neural networks such as cellular [20], binary [21], and probabilistic [22] neural networks could be considered as a swarm (network) of very simple agents (nodes) that use simple interaction principles to perform complex tasks. Several of these structures have the property of universal function approximation. Also, the theory of Boolean control networks [23] is a powerful tool for modeling, design, analysis, and simulation of cellular networks.

Some theoretical works have been performed on modeling and analysis of swarm dynamics, stability, performance and optimization [24]. Inspired by natural swarms, various computational algorithms have been proposed on swarm intelligence [25]. In case of the mentioned general methods such as in neural networks, they often require fixed patterns of interaction and interconnection among agents that are beyond the capabilities of today's nanotechnology and limitations/difficulties of nanoscale world. The goal here is to find a general swarm computing framework with a low complexity to perform complex computations in a distributed manner at nanoscale by a swarm of very simple nanometer-sized agents.

In this paper, we aim to address this problem by proposing a novel concept of swarm computing by basis agents (BAs). BA is an abstract model for a nanorobot that can compute a very simple basis function called B-function. The hardware complexity of BA is low such that its manufacturing is not far from today's technology. It is mathematically shown that a swarm of BAs is a universal function approximator and can be easily programmed to compute any arbitrary function through distributed swarm computing only by adjusting the concentration of BAs in the environment. This form of programmability is a key advantage for nanomedical applications, where physicians can then program the swarm of nanoscale BAs from outside the human body by adjusting their concentration in blood.

This paper is organized as follows. In Section 2, the concept of BA and the universal function approximation property of B-functions are explained. A specific structure for BA is proposed here with a focus on nanomedical applications. Computational accuracy and programmability of a swarm of BAs are then analytically demonstrated in Section 3. A simulation example is also provided here to illustrate the utility of the approach in this section. Finally, Section 4 concludes the paper.

## II. BASIS AGENT (BA)

Consider a given $n$-variable function $f: U \subset R^n \to R$, where $U = [a_1, b_1] \times [a_2, b_2] \times ... \times [a_n, b_n]$, $a_i$ and $b_i$ are real constants, and $a_i < b_i$ for $i = 1, ..., n$. Assume that each nanorobot can continuously sense input signals $u_1(t), u_2(t), ..., u_n(t)$ from the environment. The goal is that a swarm of nanorobots compute the value of $v(t) = f(u_1(t), u_2(t), ..., u_n(t))$ with respect to the sensed input signals at every time instant $t$. For example in nanomedical applications, the input signals could be the chemical concentrations of $n$ different species and $v(t)$ can be the concentration of drug in the aqueous environment of a diseased tissue. In such applications, the task of a swarm of nanorobots can be controlling the concentration of drug in the environment with respect to the measured input signals according to pattern $f$ defined by medical experts. Usually, a given function $f$ is more complex than it can be implemented on a single nanorobot. This section aims to show how a swarm of very simple nanoscale BAs can compute an accurate approximation of a given function $f$ at nanoscale.

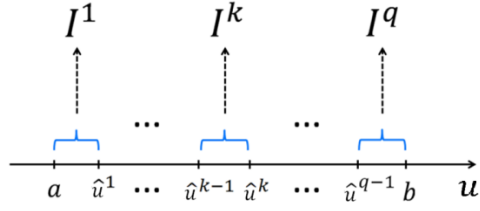

Fig.1. The domain of input $u$ is partitioned into $q$ non-overlapping intervals $I^k$ for $k = 1 : q$

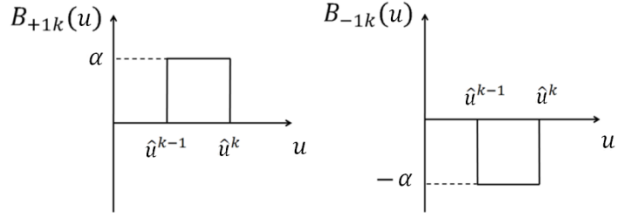

Fig.2. One-variable $B$-functions of types "$+1k$" and "$-1k$"

Assume that the domain of the $i^{th}$ input $u_i$ is partitioned into $q_i$ non-overlapping intervals $I_i^{k_i}$, where $I_i^1 = [a_i, \hat{u}_i^1)$, $I_i^{k_i} = [\hat{u}_i^{k_i-1}, \hat{u}_i^{k_i})$ for $k_i = 2, \ldots, q_i - 1$, and $I_i^{q_i} = [\hat{u}_i^{q_i-1}, b_i]$. Except $\hat{u}_i^0 = a_i$ and $\hat{u}_i^{q_i} = b_i$, the other $\hat{u}_i^{k_i}$ can be arbitrarily defined such that $\hat{u}_i^{k_i-1} < \hat{u}_i^{k_i}$ for $k_i = 1, \ldots, q_i$. As an illustration, Fig.1 shows how the domain of input $u$ (for a single-variable function) is partitioned into $q$ non-overlapping intervals $I^k$ for $k = 1 : q$. Let us define the $B$-function of type "$sk_1 k_2 \ldots k_n$" as follows:

$$B_{sk_1 k_2 \ldots k_n}(u_1, u_2, \ldots, u_n) = \begin{cases} (s)\alpha & u_1 \in I_1^{k_1} \text{ and } u_2 \in I_2^{k_2} \ \ldots \ \text{and } u_n \in I_n^{k_n} \\ 0 & \text{otherwise} \end{cases} \quad (1)$$

where $s$ is either $+1$ or $-1$, and $\alpha$ is an arbitrary positive constant. Fig.2 depicts one-variable $B$-functions of types "$+1k$" and "$-1k$". Also, the coefficient of the $B$-function of type "$sk_1 k_2 \ldots k_n$ is defined as:

$$C_{sk_1 k_2 \ldots k_n}^f = \begin{cases} \frac{R}{\alpha} |F_{k_1 k_2 \ldots k_n}| & s = sgn(F_{k_1 k_2 \ldots k_n}) \\ 0 & \text{otherwise} \end{cases} \quad (2)$$

where $F_{k_1 k_2 \ldots k_n} = f((\hat{u}_1^{k_1-1} + \hat{u}_1^{k_1})/2, \ldots, (\hat{u}_n^{k_n-1} + \hat{u}_n^{k_n})/2)$ and $R$ is an arbitrary positive constant. Eq.2 represents that $C_{sk_1 k_2 \ldots k_n}^f$ is proportional to the absolute amplitude of $f$ at the middle point of $I_1^{k_1} \times I_2^{k_2} \ldots \times I_n^{k_n}$.

**Theorem 1** (Universal Function Approximation Property). Suppose that the input universe of discourse $U$ is a compact set in $R^n$. Let $B_{sk_1 k_2 \ldots k_n}$ be a $B$-function of type "$sk_1 k_2 \ldots k_n$" on U defined in Eq. 1. Then, for any given real continuous function $f : U \subset R^n \to R$ and arbitrary $\epsilon > 0$, there exist an integer $N = 2 \prod_{i=1}^n q_i$ and real constants $C_{sk_1 k_2 \ldots k_n}^f$ defined in Eq. 2, such that:

$$\left\| f - \sum_{k_1=1}^{q_1} \ldots \sum_{k_n=1}^{q_n} \sum_{s=\{-1,+1\}} \frac{1}{R} C_{sk_1 k_2 \ldots k_n}^f B_{sk_1 k_2 \ldots k_n} \right\|_\infty < \epsilon$$

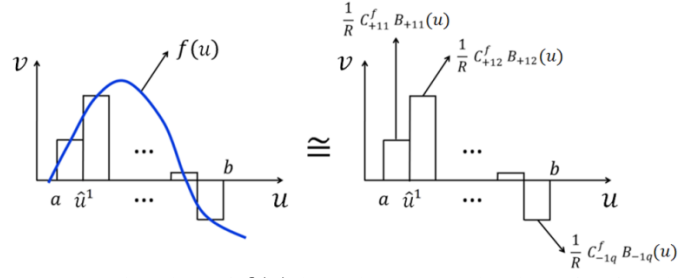

Fig.3. Approximation of $f(u)$ by a linear combination of $B$-functions

**Proof.** According to the sifting property of Dirac delta function [26] and as displayed in Fig.3 (for one-variable $f$), for every $[u_1 \, u_2 \, \dots \, u_n] \in U$, $f(u_1, u_2, \dots, u_n)$ can be approximated by a linear combination of $N$ types of $B$-functions with the coefficients $\frac{1}{R} C^f_{s k_1 k_2 \dots k_n}$ as follows:

$$f(u_1, u_2, \dots, u_n) \cong \sum_{k_1=1}^{q_1} \dots \sum_{k_n=1}^{q_n} \sum_{s=\{-1,+1\}} \frac{1}{R} C^f_{s k_1 k_2 \dots k_n} B_{s k_1 k_2 \dots k_n}(u_1, u_2, \dots, u_n) \tag{3}$$

where $N = 2 \prod_{i=1}^{n} q_i$. The approximation accuracy of Eq.3 is improved by increasing $q_i$ such that approximation error tends to zero when all $q_i$ (for $i = 1, \dots, n$) approach infinity [26]:

$$\lim_{q_1 \to \infty, \dots, q_n \to \infty} \left\| f - \sum_{k_1=1}^{q_1} \dots \sum_{k_n=1}^{q_n} \sum_{s=\{-1,+1\}} \frac{1}{R} C^f_{s k_1 \dots k_n} B_{s k_1 \dots k_n} \right\|_\infty \to 0$$

So, Eq.3 demonstrates the ability of $B$-functions as universal function approximator.∎

Inspired by the concept of basis function, *Basis Agent* (BA) of type "$s k_1 k_2 \dots k_n$" is defined as a nanorobot that can sense the input signals $u_1, u_2, \dots, u_n$ from the environment, compute the $B$-function of type "$s k_1 k_2 \dots k_n$", and deliver the computed value as the output of BA to the environment. Also, the concentration of BA of type "$s k_1 k_2 \dots k_n$" in the environment is set to $C^f_{s k_1 k_2 \dots k_n}$. According to Theorem 1, it can be concluded that a swarm of $N$ types of BAs ($N$ is defined in Theorem 1) offer the capability of universal function approximation. This is illustrated more in further detail in Section 3.

### A. A Proposed Nano-Structure for BA

The above mathematical description of BA can be implemented in different forms depending on the problem type and technology. Without loss of generality, this paper aims to focus on nanomedical applications. In the rest of this study, we assume that the swarm of BAs is working in the aqueous environment of a living tissue. The input signals $u_1(t), u_2(t), \dots, u_n(t)$ are chemical concentrations of $n$ different species in the environment that are continuously sensed by every BA. The output signal of a BA of type "$s k_1 k_2 \dots k_n$" is the release (if $s = +1$) or uptake (if $s = -1$) rate of drug molecules. Fig.4 displays our proposed structure for a BA of type "$s k_1 k_2 \dots k_n$". The pump is pumping drug molecules with constant flow rate $(s)\alpha$. As shown in Fig.5, the flow direction of pump is defined by the sign of $s$, i.e. release if $s = +1$, and uptake if $s = -1$. For each input $u_i$, BA has two on-off valves in series that are controllable by the amplitude of $u_i$. The behavior of these valves is illustrated in Fig.6.

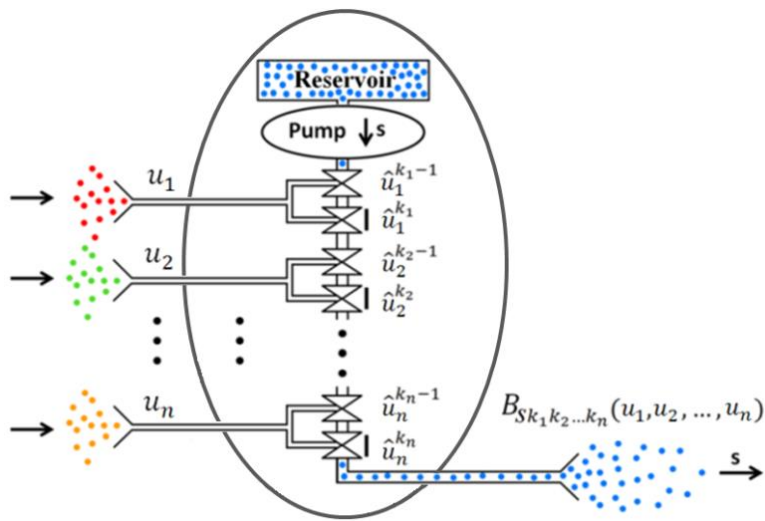

Fig.4. The structure of a BA of type "$sk_1k_2 \ldots k_n$". The flow direction of pump and output in this figure is related to $s = +1$. If $s = -1$, the flow direction of pump and output is reversed (see Fig.5).

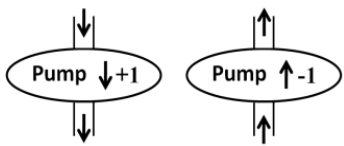

Fig.5. The behavior of pump in Fig.4 for $s = +1$ (left) and $s = -1$ (right). The arrow shows the flow direction.

## B. Experimental Feasibility

A basis agent consists of molecular concentration sensors, nanoscale computing unit, and nanometer-sized controllable valve. In recent years, many remarkable experimental works have been performed on the construction of computing units, sensors and actuators at nanoscale. Advanced nanostructures as well as biotechnology-based nanoscale products have been widely used for designing nanosensors such as functionalized nanotubes for in-vivo pharmacological measurements of a chemotherapy drug [27], dual-ligand functionalized carbon nanodots for cellular targeted imaging [28], smart multifunctional nanoparticles for drug delivery [29], mesoporous silica nanoparticles for sensitive detection of a psychedelic drug [30], DNA-based reversible nanosensor [31], and aptamer-based bio-nanosensors [32]. Similarly, different technologies have been employed for constructing nanoactuators such as mechanosensitive channel [33], mesoporous silica nanoparticles [34], stimuli-responsive nanovalves [35], optogenetics [36], and DNA-based nanoactuators [37].

The notion of realizing molecular computers for biomedical applications has been recognized as a fundamental problem in the literature [38] and different techniques have been proposed for implementation of programmable circuits and logic programs at molecular [39], intracellular [40], and intercellular [41] levels. Also, some researches have been performed on the construction of memory [42-43] and automata [44-45] for molecular and cellular environments. In order to design nanoscale computing units for in-vivo applications, DNA technology is one of the most common approaches. Some examples of DNA-based biocomputing units include synthetic molecules for smart information processing [46], molecular system for cancer diagnosis [47], fuzzy, Boolean and universal logic gates [48-49], programmable chemical controllers [50], and complex computing circuits made from DNA-based building blocks [51-52]. Generally, these previous works are promising enough to demonstrate that smaller nanoscale components will be manufactured in near future and manufacturing of BA will be experimentally feasible.

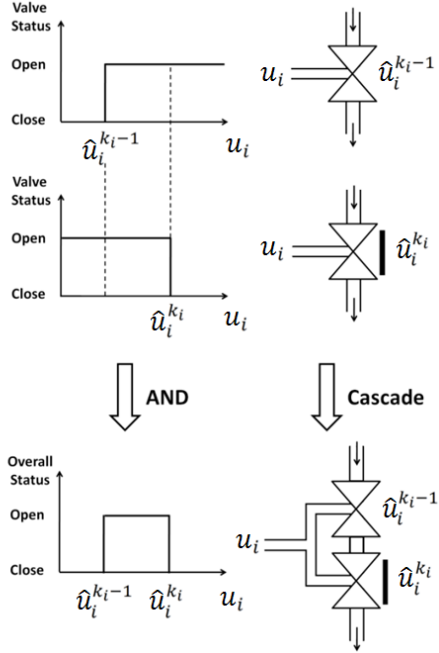

Fig.6. The behavior of valves in Fig.4. The flow direction of valves is related to $s = +1$.

## III. SWARM COMPUTING BY BASIS AGENTS

The task of a swarm of BAs is to automatically control the concentration of drug, $v(t)$, in the environment with respect to the measured input signals $u_1(t), \dots, u_n(t)$ according to pattern $v(t) = f(u_1(t), \dots, u_n(t))$ defined by medical experts.

This pattern is allowed to be changed over time. In a unit volume of space of arbitrarily small size, the governing equation between input signals and $v(t)$ is:

$$\dot{v}(t) = -R\, v(t) + \sum_{k_1=1}^{q_1} \dots \sum_{k_n=1}^{q_n} \sum_{s=\{-1,+1\}} C^f_{sk_1 \dots k_n} B_{sk_1 \dots k_n}\big(u_1(t), \dots, u_n(t)\big) \qquad (4)$$

where $B_{sk_1 k_2 \dots k_n}(u_1(t), \dots, u_n(t))$ is the drug release or uptake (depending on the sign of $s$) rate by each BA of type $sk_1 k_2 \dots k_n$, and $C^f_{sk_1 k_2 \dots k_n}$ is the concentration of BA of type $sk_1 k_2 \dots k_n$ in the environment according to Eq.2. Fig.7 illustrates the concept of swarm computing by BAs for a given single-variable function in a small volume of an aqueous environment. In this figure, red and blue dots stand for the particles of the input signal ($u$) and drug ($v$), respectively. Also, according to the definition of the pattern $f(u)$ in this figure, when the input $u$ is $u^* \in I^2$, only the BAs of type "+12" are activated which are highlighted in green.

In Eq.4, concentration is defined as the number of particles in a unit volume of space. $R$ is the elimination rate constant for drug that is a positive constant. It is important to note that such a clearance mechanism (not necessarily linear) is useful to automatically damp the effect of past inputs. If this clearance mechanism does not exist, $v(t)$ becomes an accumulator. In living environments, such clearance mechanisms usually exist due to different causes such as reaction between drug and other species as well as drug uptake by cells. In this paper, we assume that the concentration of BAs in the environment can be controlled and quickly changed by an external mechanism, which the details of this mechanism is not in the scope of this study. Also, the reservoir of BA contains sufficient drug molecules and will not become empty during the lifetime of BA.

### A. Computational Accuracy

Assume that the task of a swarm of BAs is to compute $f_1(u_1(t), ..., u_n(t))$ for $t \in [t_0, t_2]$. According to Eq.2, the concentration of a BA of type $sk_1 ... k_n$ in the environment should be set to $C_{sk_1 k_2 ... k_n}^{f_1} = \frac{R}{u} \left| f_1((\hat{u}_1^{k_1-1} + \hat{u}_1^{k_1})/2, ..., (\hat{u}_n^{k_n-1} + \hat{u}_n^{k_n})/2) \right|$. Also, for $t \in [t_0, t_1]$, $u_1(t) \in I_1^{\bar{k}_1}$, $u_2(t) \in I_2^{\bar{k}_2}$ ..., $u_n(t) \in I_n^{\bar{k}_n}$ and consequently the desired value of $v(t)$ is $f_1((\hat{u}_1^{\bar{k}_1-1} + \hat{u}_1^{\bar{k}_1})/2, ..., (\hat{u}_n^{\bar{k}_n-1} + \hat{u}_n^{\bar{k}_n})/2)$.

**Theorem 2** (Computational Accuracy). A swarm of BAs can precisely compute any given function in steady state when the transient computational error $e(t)$ is bounded, that is, there exist a finite $M > 0$ such that

$$\|e\|_\infty \le M$$

and

$$\lim_{t \to \infty} e(t) \to 0.$$

**Proof**. According to the definitions of $C_{sk_1 k_2 ... k_n}^{f_1}$ and $B_{sk_1 k_2 ... k_n}(u_1(t), ..., u_n(t))$, $C_{\bar{s}\bar{k}_1 \bar{k}_2 ... \bar{k}_n}^{f_1} B_{\bar{s}\bar{k}_1 \bar{k}_2 ... \bar{k}_n}(u_1(t), ..., u_n(t))$ is only nonzero and equal to $(\bar{s})\alpha \, C_{\bar{s}\bar{k}_1 \bar{k}_2 ... \bar{k}_n}^{f_1}$ for $t \in [t_0, t_1]$, where $\bar{s} = sgn(f_1((\hat{u}_1^{\bar{k}_1-1} + \hat{u}_1^{\bar{k}_1})/2, ..., (\hat{u}_n^{\bar{k}_n-1} + \hat{u}_n^{\bar{k}_n})/2))$.

As a result, for $t \in [t_0, t_1]$, the second term of the right side of Eq.4 is simplified to:

$$\sum_{k_1=1}^{q_1} ... \sum_{k_n=1}^{q_n} \sum_{s=\{-1,+1\}} C_{sk_1 k_2 ... k_n}^{f_1} B_{sk_1 k_2 ... k_n}(u_1(t), ..., u_n(t)) =$$

$$C_{\bar{s}\bar{k}_1 \bar{k}_2 ... \bar{k}_n}^{f_1} B_{\bar{s}\bar{k}_1 \bar{k}_2 ... \bar{k}_n}(u_1(t), ..., u_n(t)) = R \, f_1((\hat{u}_1^{\bar{k}_1-1} + \hat{u}_1^{\bar{k}_1})/2, ..., (\hat{u}_n^{\bar{k}_n-1} + \hat{u}_n^{\bar{k}_n})/2) \tag{5}$$

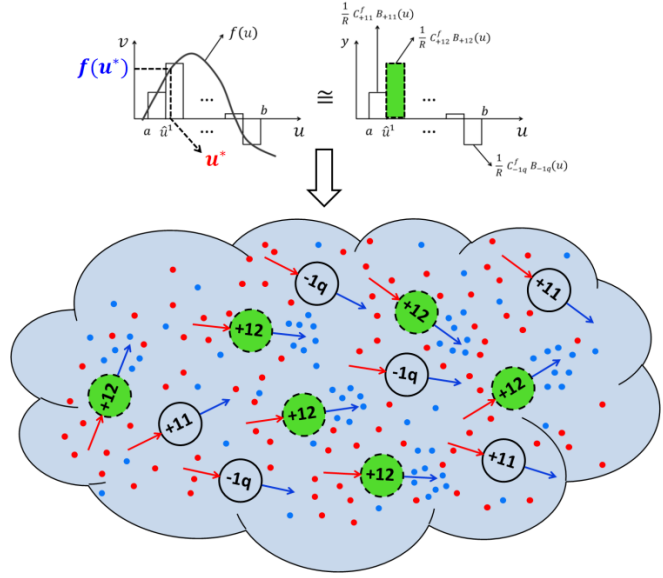

Fig.7. This figure shows the concept of swarm computing by BAs for a given single-variable function in a small volume of an aqueous environment. When the input $u$ is $u^* \in I^2$, only the BAs of type "+12" are activated (highlighted with dotted border).

By substituting Eq.5 in Eq.4 and solving the obtained differential equation, $v(t)$ is calculated as follows for $t \in [t_0, t_1]$:

$$v(t) = v(t_0)e^{-R(t-t_0)} + f_1((\hat{u}_1^{\bar{k}_1-1} + \hat{u}_1^{\bar{k}_1})/2, \dots, (\hat{u}_n^{\bar{k}_n-1} + \hat{u}_n^{\bar{k}_n})/2)(1 - e^{-R(t-t_0)}) \qquad (6)$$

The steady state value and time constant of $v(t)$ in Eq.6 are $v_{ss} = f_1((\hat{u}_1^{\bar{k}_1-1} + \hat{u}_1^{\bar{k}_1})/2, \dots, (\hat{u}_n^{\bar{k}_n-1} + \hat{u}_n^{\bar{k}_n})/2)$ and $\tau_v = \frac{1}{R}$, respectively. The transient computational error $e(t)$ is defined as follows for $t \in [t_0, t_1]$:

$$e(t) = v_{ss} - v(t) = (f_1((\hat{u}_1^{\bar{k}_1-1} + \hat{u}_1^{\bar{k}_1})/2, \dots, (\hat{u}_n^{\bar{k}_n-1} + \hat{u}_n^{\bar{k}_n})/2) - v(t_0)) \, e^{-R(t-t_0)} \qquad (7)$$

According to Equations 6 and 7, after $5\tau_v$ units of time, $e(t)$ is damped and $v(t)$ reaches 99.3% of its steady state value. It is important to note that the steady state value of $v(t)$ in Eq.6 is equal to $f_1((\hat{u}_1^{\bar{k}_1-1} + \hat{u}_1^{\bar{k}_1})/2, \dots, (\hat{u}_n^{\bar{k}_n-1} + \hat{u}_n^{\bar{k}_n})/2)$ which demonstrates that the swarm of BAs could precisely compute the desired value of $v(t)$ in steady state. Eq.7 shows that:

$$\|e\|_\infty \le \left| (f_1((\hat{u}_1^{\bar{k}_1-1} + \hat{u}_1^{\bar{k}_1})/2, \dots, (\hat{u}_n^{\bar{k}_n-1} + \hat{u}_n^{\bar{k}_n})/2) - v(t_0)) \right|$$

and $\lim_{t \to \infty} e(t) \to 0. \blacksquare$

It should be noted that since in real-world applications, the number and the types of basis agents are finite, there always exists a mathematical-approximation error between the desired function and its approximation by basis agents according to Theorem 1. Thus, although the transient-computational error disappears as shown in Theorem 3, the mathematical-approximation error may still remain.

Now suppose that, for $t \in (t_1, t_2]$, $u_1(t) \in I_1^{\bar{k}_1}, u_2(t) \in I_2^{\bar{k}_2} \dots, u_n(t) \in I_n^{\bar{k}_n}$ and consequently the desired value of $v(t)$ is $f_1((\hat{u}_1^{\bar{k}_1 - 1} + \hat{u}_1^{\bar{k}_1})/2, \dots, (\hat{u}_n^{\bar{k}_n - 1} + \hat{u}_n^{\bar{k}_n})/2)$. In this time interval, only $C_{\bar{s}\bar{k}_1\bar{k}_2\dots\bar{k}_n}^{f_1} B_{\bar{s}\bar{k}_1\bar{k}_2\dots\bar{k}_n}(u_1(t), \dots, u_n(t))$ is nonzero and the outputs of the all other types of BAs including type $\bar{s}\bar{k}_1\bar{k}_2\dots\bar{k}_n$ are made zero. If $(t_1 - t_0) \gg 5\tau_v$, $y(t_1) \cong f_1((\hat{u}_1^{\bar{k}_1 - 1} + \hat{u}_1^{\bar{k}_1})/2, \dots, (\hat{u}_n^{\bar{k}_n - 1} + \hat{u}_n^{\bar{k}_n})/2)$. Similar to Eq.6, $v(t)$ is computed as follows for $t \in (t_1, t_2]$:

$$v(t) = f_1((\hat{u}_1^{\bar{k}_1 - 1} + \hat{u}_1^{\bar{k}_1})/2, \dots, (\hat{u}_n^{\bar{k}_n - 1} + \hat{u}_n^{\bar{k}_n})/2)e^{-R(t - t_1)} + f_1((\hat{u}_1^{\bar{k}_1 - 1} + \hat{u}_1^{\bar{k}_1})/2, \dots, (\hat{u}_n^{\bar{k}_n - 1} + \hat{u}_n^{\bar{k}_n})/2)(1 - e^{-R(t - t_1)})$$

(8)

According to this equation, the effect of $f_1((\hat{u}_1^{\bar{k}_1 - 1} + \hat{u}_1^{\bar{k}_1})/2, \dots, (\hat{u}_n^{\bar{k}_n - 1} + \hat{u}_n^{\bar{k}_n})/2)$ is damped over time and $v(t)$ reaches 99.3% of $f_1((\hat{u}_1^{\bar{k}_1 - 1} + \hat{u}_1^{\bar{k}_1})/2, \dots, (\hat{u}_n^{\bar{k}_n - 1} + \hat{u}_n^{\bar{k}_n})/2)$ after $5\tau_v$ units of time. Equations 6-8 demonstrate that the swarm of BAs can compute $f_1(u_1(t), \dots, u_n(t))$ over all the input space.

## B. Programmability

Assume that, for $t \in (t_2, t_3]$, the inputs are similar to $t \in (t_1, t_2]$ (i.e. $u_1(t) \in I_1^{\bar{k}_1}, u_2(t) \in I_2^{\bar{k}_2} \dots, u_n(t) \in I_n^{\bar{k}_n}$). In this time interval, we aim to program the swarm of BAs to compute $f_2(u_1, \dots, u_n)$ instead of $f_1(u_1, \dots, u_n)$.

**Theorem 3** (Programmability). A swarm of BAs can be programmed to compute $f_2(u_1, \dots, u_n)$ instead of $f_1(u_1, \dots, u_n)$ by changing the concentrations of BAs from $C_{sk_1k_2\dots k_n}^{f_1}$ to $C_{sk_1k_2\dots k_n}^{f_2} = \frac{R}{u}|f_2((\hat{u}_1^{k_1 - 1} + \hat{u}_1^{k_1})/2, \dots, (\hat{u}_n^{k_n - 1} + \hat{u}_n^{k_n})/2)|$ in the environment.

**Proof.** Since the goal is to program the swarm of BAs to compute $f_2(u_1, \dots, u_n)$ instead of $f_1(u_1, \dots, u_n)$, we change the concentrations of all types of BAs from $C_{sk_1k_2\dots k_n}^{f_1}$ to $C_{sk_1k_2\dots k_n}^{f_2} = \frac{R}{u}|f_2((\hat{u}_1^{k_1 - 1} + \hat{u}_1^{k_1})/2, \dots, (\hat{u}_n^{k_n - 1} + \hat{u}_n^{k_n})/2)|$. According to Eq.5:

$$\sum_{k_1 = 1}^{q_1} \dots \sum_{k_n = 1}^{q_n} \sum_{s = \{-1, +1\}} C_{sk_1\dots k_n}^{f_2} B_{sk_1\dots k_n}(u_1(t), \dots, u_n(t)) =$$

$$C_{\bar{s}\bar{k}_1\bar{k}_2\dots\bar{k}_n}^{f_2} B_{\bar{s}\bar{k}_1\bar{k}_2\dots\bar{k}_n}(u_1(t), \dots, u_n(t)) = R\, f_2((\hat{u}_1^{\bar{k}_1 - 1} + \hat{u}_1^{\bar{k}_1})/2, \dots, (\hat{u}_n^{\bar{k}_n - 1} + \hat{u}_n^{\bar{k}_n})/2)$$

(9)

If $(t_2 - t_1) \gg 5\tau_v$, $v(t_2) \cong f_1((\hat{u}_1^{\bar{k}_1 - 1} + \hat{u}_1^{\bar{k}_1})/2, \dots, (\hat{u}_n^{\bar{k}_n - 1} + \hat{u}_n^{\bar{k}_n})/2)$. Similar to Eq.6, $v(t)$ is obtained as follows for $t \in (t_2, t_3]$:

$$v(t) = f_1((\hat{u}_1^{\bar{k}_1 - 1} + \hat{u}_1^{\bar{k}_1})/2, \dots, (\hat{u}_n^{\bar{k}_n - 1} + \hat{u}_n^{\bar{k}_n})/2)\, e^{-R(t - t_2)} + f_2((\hat{u}_1^{\bar{k}_1 - 1} + \hat{u}_1^{\bar{k}_1})/2, \dots, (\hat{u}_n^{\bar{k}_n - 1} + \hat{u}_n^{\bar{k}_n})/2)(1 - e^{-R(t - t_2)})$$

(10)

As expected, the steady state value of $v(t)$ in Eq.10 is $f_2((\hat{u}_1^{\overline{k}_1-1} + \hat{u}_1^{\overline{k}_1})/2, \dots, (\hat{u}_n^{\overline{k}_n-1} + \hat{u}_n^{\overline{k}_n})/2)$ which means that after $5\tau_v$ units of time, the swarm of BAs can forget $f_1(u_1, \dots, u_n)$ and compute $f_2(u_1, \dots, u_n)$ instead of $f_1(u_1, \dots, u_n)$. This demonstrates that the swarm of BAs can be easily programmed by adjusting the concentrations of BAs in the environment.∎

In this paper, for ease of implementation, we assume that the output of BA is constant through time. By this assumption, $R$ is the only parameter that affects the speed of computations and programmability. It should be noted that $R$ is an environmental parameter and usually it cannot be manipulated by designer. In cases where $R$ is very small ($\tau_v$ is very large), one way to improve the speed of computations and programmability is to design BAs with time-varying (non-constant) output, while this will increase the hardware complexity of BA. In this design, when a BA is activated by inputs, first the amplitude of its output should be larger than $\alpha$, and then it must decrease through time until it tends to $\alpha$. This causes that $v(t)$ reaches its steady state value sooner than $\tau_v$.

### C. Minimal Design

Finding the minimum number of BAs types is practically important since it considerably reduces the hardware complexity of the swarm. Given specific approximation accuracy, the following theorem can be used to determine the near-minimal number of required BAs types to reach the desired accuracy.

**Theorem 4** (Near-minimal Number of Basis Agents). Assume that $v_f$ is designed according to Theorem 1 to approximate a given function $v = f(u_1, \dots, u_n)$ on $U$ with a desired approximation accuracy of $\varepsilon$ such that:

$$\forall [u_1 \dots u_n] \in U: \ |v - v_f| \leq \varepsilon$$

If the $f(u_1, \dots, u_n)$ is continuously differentiable on $U$, then the near-minimal number of the types of BAs, i.e. $N_{min}$, for reaching this accuracy is:

$$N_{min} = \min_{h_1, \dots, h_n} 2 \prod_{i=1}^{n} \left\lceil \frac{(b_i - a_i)}{h_i} \right\rceil$$

$$\text{s.t.} \begin{cases} \sum_{i=1}^{n} \left\| \frac{\partial f}{\partial u_i} \right\|_{\infty} h_i \leq \varepsilon \\ 0 < h_i \leq (b_i - a_i) \qquad \text{for } i = 1, \dots, n \end{cases}$$

where $a_i$ and $b_i$ are the lower and upper bounds of $u_i$, $\left\| \frac{\partial f}{\partial u_i} \right\|_{\infty}$ is the infinity norm of $\frac{\partial f}{\partial u_i}$ on $U$, and $\left\lceil \frac{(b_i - a_i)}{h_i} \right\rceil$ is the ceiling of $\frac{(b_i - a_i)}{h_i}$ that is the smallest integer not less than $\frac{(b_i - a_i)}{h_i}$.

**Proof:** Given $v = f(u_1, \dots, u_n)$, assume that $v_f$ is designed according to Theorem 1. As it has been proven in [53], if the $f(u_1, \dots, u_n)$ is continuously differentiable on $U$, then:

$\forall [u_1 \dots u_n] \in U: \ \left| v - v_f \right| \le \sum_{i=1}^{n} \left\| \frac{\partial f}{\partial u_i} \right\|_{\infty} h_i$

where $h_i = \max\limits_{k_i=0,\dots,q_i-1} \left| \hat{u}_i^{k_i+1} - \hat{u}_i^{k_i} \right|$, for $i = 1, \dots, n$. As a result, to ensure that $\forall [u_1 \dots u_n] \in U$: $\left| v - v_f \right| \le \varepsilon$, it is enough to ensure that $\sum_{i=1}^{n} \left\| \frac{\partial f}{\partial u_i} \right\|_{\infty} h_i \le \varepsilon$ for continuously differentiable functions. As described earlier, the number of the types of BAs is equal to $N = 2 \prod_{i=1}^{n} q_i$. Regarding the definition of $h_i$ and $q_i$, it is obvious that $q_i = \left\lceil \frac{(b_i - a_i)}{h_i} \right\rceil$ and therefore the number of the types of BAs is equal to $2 \prod_{i=1}^{n} \left\lceil \frac{(b_i - a_i)}{h_i} \right\rceil$. Consequently, the goal is to find the optimal values of $h_1, \dots, h_n$ such that the above-mentioned constraint is satisfied and the number of the types of BAs is minimized. Solving the nonlinear programming problem mentioned in this theorem gives the near-minimal number of the types of BAs which can approximate the given function with the desired accuracy. ∎

It should be noted that since the above-mentioned constraints in the proof are the upper bounds of $\left| v - v_f \right|$, this theorem just gives a near-minimal number of the types of BAs. This means that it is probable that the given function can be approximated with the desired accuracy by a less number of the types of BAs. To find the minimal number of the types of BAs, we recommend using population-based global optimization methods such as genetic algorithms. Given a specific problem, these optimization methods can efficiently find the minimal number of the types of BAs as well as the optimal values of the structural parameters of BAs such as the number of sensors (inputs) and the values of controller unit parameters.

### D. Simulation Example

In this section, the performance of the proposed approach is illustrated by a simulation example. Two test input signals, i.e. $u(t)$, are considered including ramp and step inputs. These signals are displayed in Fig.12-a, where their amplitude varies between 0 and 1. We aim to program a swarm of BAs to compute the following time-varying function:

$$v(t) = f(u(t), t) = \begin{cases} f_1\big(u(t)\big) = u(t)^2 & 0 \le t < 200 \\ f_2\big(u(t)\big) = \sin(3u(t)) & 200 \le t < 400 \\ f_3\big(u(t)\big) = e^{-2u(t)} & 400 \le t \le 600 \end{cases} \tag{11}$$

Since the amplitude of $f(u)$ in Eq.11 is always positive over $[0, 1]$, the concentration of BAs of type "$-1k$" is zero for all $k$. In this simulation, the domain of $u$, i.e. $[0, 1]$, is partitioned into 10 equal intervals ($q = 10$). Also, $\hat{u}^k = 0.1k$ for $k = 0, 1, \dots, 10$, $v(0) = 0$, and $\alpha = R = 1$. Fig.8 indicates $f_1(u)$, $f_2(u)$ and $f_3(u)$ and their approximations by linear combination of 10 $B$-functions according to Eq.3. Fig.9 shows the structure of a BA of type $+1k$ used in this example as a special case of Fig.4. For each function, the concentration of each type of BAs has been calculated by Eq.2 and illustrated in Fig.10. To program the swarm, the concentration of each type of BA must be changed over time according to Fig.11. In this example, it is assumed that the time constant of changes in concentration (by an external mechanism) is much smaller than the time constant of $v(t)$ in Eq.6. Hence, the process of changes in concentration is ideally considered as an instantaneous process as shown in Fig.11. Fig.12-b and c compare the computed value of $v(t)$ by the swarm of BAs vs. its desired value for the test input $u(t)$, where the desired value of $v(t)$ is defined as the approximated value of $f(u(t), t)$ by linear combination of 10 $B$-functions according to Fig.8. The transient computational error (Eq.7) is represented in Fig12-d. As it is proven in Theorem 2, this figure shows that the computational error is bounded and it tends to zero in steady state. The mean absolute of transient computational error (MAE) is $0.0121$ ($1.21\%$ of the full-scale) and $0.0026$ ($0.26\%$ of the full-scale) for ramp and step test inputs, respectively. Fig.12 demonstrates that the swarm of BAs can accurately compute the desired functions, and also it can be easily programmed only by adjusting the concentrations of BAs over time.

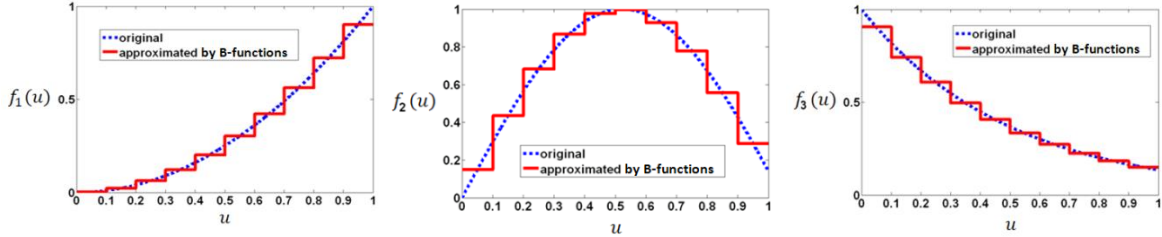

Fig.8. The functions $f_1(u)$, $f_2(u)$ and $f_3(u)$ and their approximations by linear combination of 10 $B$-functions according to Eq.3.

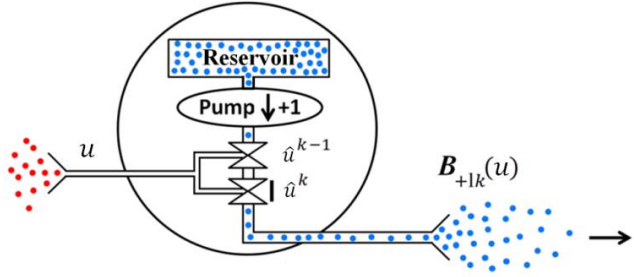

Fig.9. The structure of a one-variable BA of type "$+1k$" used in the simulation example.

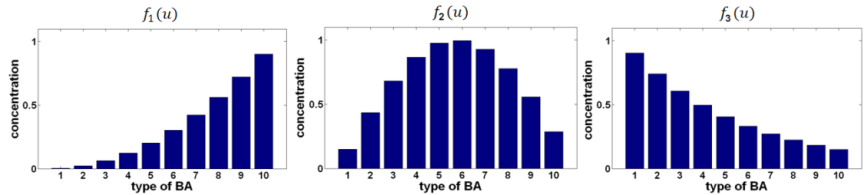

Fig.10. The concentration of BAs of type "$+1k$" for $f_1(u)$, $f_2(u)$ and $f_3(u)$. The horizontal axis indicates index $k$.

## IV. CONCLUSION

This paper proposes a novel approach towards realization of programmable computing on a swarm of very simple nanorobots in a distributed manner. The proposed method is based on the idea of swarm computing by basis agents (BAs). The hardware complexity of each BA is low. It is mathematically proven that a swarm of BAs has the universal function approximation property and also it can be easily programmed to accurately compute desired functions only by adjusting the concentrations of BAs in the environment over time. This form of programmability is a key advantage for nanomedical applications, where physicians can program the swarm of nanoscale BAs from outside the human body by adjusting their concentration in blood. Simplicity, universality, computational accuracy and programmability of swarm computing by the proposed BAs demonstrate that this method is a promising approach towards realization of programmable computers on nanorobots.

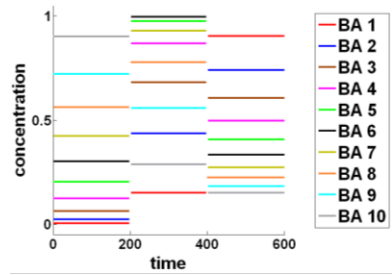

Fig.11. The swarm of BAs can be easily programmed by adjusting the concentration of each type of BAs over time as displayed in this figure.

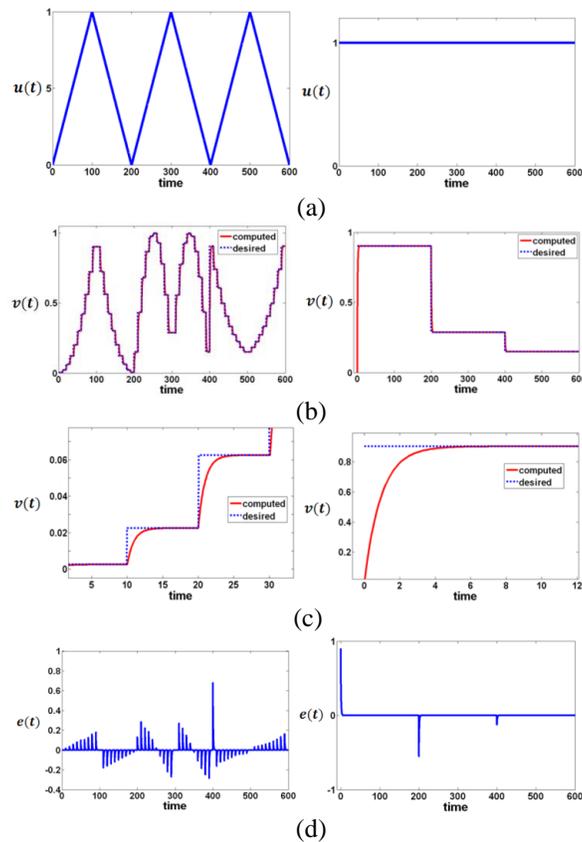

Fig.12. Simulation results for ramp (left plots) and step (right plots) test inputs. a) Test input $u(t)$, b) Computed value of $v(t)$ by the swarm of BAs vs. its desired value for the test input $u(t)$, where the desired value of $v(t)$ is defined as the approximated value of $f(u(t), t)$ by linear combination of 10 $B$-functions according to Fig.8., c) A zoomed-in plot of Figure b, d) Transient computational error.